\documentclass[%
 reprint,
 amsmath,amssymb,
 aps,
]{revtex4-2}

\usepackage{graphicx}
\usepackage{dcolumn}
\usepackage{bm}
\usepackage{xcolor}
\usepackage{subfiles}
\usepackage{wrapfig}
\usepackage{xr}
\newcommand{\SK}[1]{\textcolor{black}{{#1}}}

\begin{document}

\preprint{APS/123-QED}

\title{Inverse Bauschinger to Bauschinger Crossover under Steady Shear in Amorphous Solids}

\author{Rashmi Priya$^1$}
 \email{rpriya@tifrh.res.in}
\author{Smarajit Karmakar$^1$} 
 \email{smarajit@tifrh.res.in}
\affiliation{ $^1$ Tata Institute of Fundamental Research, 36/P, Gopanpally Village, Serilingampally Mandal,
Ranga Reddy District, Hyderabad, India 500046}

\begin{abstract}
\SK{Directional memory in amorphous solids is commonly quantified through the Bauschinger effect, yet the observation of the inverse Bauschinger effect suggests that the sign of memory can invert, pointing to distinct underlying plastic organization. Here, we connect directional memory to the nature of yielding in steadily sheared amorphous solids. Using simulations of two-dimensional polydisperse glasses, we show that the type of directional memory (Bauschinger versus inverse Bauschinger) is jointly controlled by deformation history, strain rate, and parent temperature. We identify a critical history amplitude $\gamma_{N,\mathrm{crit}}(T_p,\dot{\gamma})$ and construct a phase diagram that delineates regimes with memory inversion from those showing only conventional Bauschinger response. Microscopically, memory inversion correlates with network-like shear-band morphology and plastic healing, whereas conventional memory is associated with persistent localization and cumulative damage. These results establish directional memory as an order parameter for a shear-rate and annealing-controlled brittle-ductile crossover and suggest that plastic healing provides a generic route to memory inversion in disordered solids.}

\end{abstract}

\maketitle


\section{Introduction\protect\\}

\SK{Amorphous solids are out-of-equilibrium materials whose yielding and failure span brittle shear localization and ductile flow \cite{Berthier_2025_YieldingPlasticity, PhysRevLett.125.168003, article, 10.1063/5.0086626, PhysRevResearch.2.023203}. These modes of failure depend on preparation (annealing), system size, and the applied strain rate, which together shape the stability of the glass and the energy-barrier landscape underlying plastic rearrangements and mechanical memory \cite{article, PhysRevMaterials.4.025603, maass2015shear}. The accumulation of irreversible rearrangements encodes deformation-induced anisotropy in the microstructure, which can later be retrieved as directional memory. Directional memory is commonly viewed as the persistence of this deformation-induced anisotropy, yet its connection to the nature of yielding remains unclear.}

\SK{Under homogeneous shear, directional memory is commonly studied in terms of the Bauschinger effect: after a pre-shear, the yield stress upon reversing the loading direction is reduced compared to reloading in the same direction, reflecting a stored structural anisotropy that encodes the prior loading direction \cite{RevModPhys.91.035002, PhysRevLett.81.3243, Bauschinger, Bouaziz_2023_Bauschinger}. In polycrystalline materials, this asymmetry is often rationalized in terms of internal backstress/kinematic hardening generated by dislocation structures, while in amorphous solids it is attributed to anisotropic distributions of local yielding thresholds and residual stresses created by prior plastic flow \cite{bouchbinder2010nonequilibrium, PhysRevLett.124.205503, Koizumi2016, PhysRevE.82.026104}. While the Bauschinger effect has long served as a hallmark of directional memory in amorphous solids, the recent discovery of the inverse Bauschinger effect indicates that directional memory can invert, pointing to a distinct underlying organization of plasticity \cite{rpriya_IBE}. The inverse Bauschinger effect has been reported in crystalline and polycrystalline solids under specific microstructural conditions, for example involving the formation of dislocation twins near grain boundaries during reverse loading, as discussed for polycrystalline aluminum wires \cite{Koizumi2016}, and, more recently, in active-doped ultrastable glasses, where it has been linked to plastic healing in shear-band networks \cite{rpriya_IBE}. This raises the question of the roles that activity and structural organisation play in memory inversion, and whether it can also emerge under steady shear in passive glasses. How does preparation and driving control memory inversion, and is it tied to a change in yielding from brittle localisation to ductile flow? Finally, what microscopic organization of plastic rearrangements distinguishes conventional and inverted directional memory?}

\SK{Mechanical yielding in amorphous solids emerges from the accumulation of irreversible rearrangements, accompanied by a marked increase in energy dissipation \cite{Yu_2024_GlassyRelaxation}. In structural glasses, yielding and failure depend sensitively on both system size and preparation. Increasing the system size promotes brittle-like yielding, often reflected in a sharp stress drop at yield and the emergence of a system-spanning shear band \cite{article, PhysRevLett.125.168003, sergueeva2005shear}. Preparation is commonly characterized by the parent temperature $T_p$, which sets the kinetic stability of the glass: lower $T_p$ corresponds to more stable, brittle-like glasses with higher activation barriers, whereas higher $T_p$ yields less stable, more ductile glasses. Well-annealed states can be accessed through various means such as slow cooling, aging, oscillatory shear, or activity-induced annealing, etc \cite{article, srikanth_finite_temp, 10.1073/pnas.2100227118, vishnu_etal, PhysRevLett.124.225502, biroli_aging, Sharma2025}.}

\SK{A particularly important limit is that of ultrastable glasses, prepared via specialized protocols such as swap Monte Carlo in simulations or physical vapor deposition (PVD) in experiments \cite{PhysRevE.63.045102, Berthier_2019, PhysRevX.7.021039, 2022NCimR..45..325R, PhysRevLett.125.085505, Ozawa2023, https://doi.org/10.1002/adma.201302700}. These methods reach deeper annealing than conventional routes and generate glasses that are mechanically, thermodynamically, and kinetically more stable than ordinary glasses, often exhibiting extremely brittle-like yielding \cite{2022NCimR..45..325R}. Beyond preparation, the strain rate strongly influences how plasticity organizes at yielding: in the athermal quasistatic (AQS) or very slow-rate limit, failure commonly localizes into a single dominant band, whereas at higher strain rates plastic activity can produce multiple shear bands that form a network \cite{PhysRevMaterials.4.025603, liu2005behavior}. Recent studies, including work on active-doped brittle-like glasses, have emphasized such rate-dependent changes in shear-band morphology and statistics \cite{rpriya2025}. However, how these distinct yielding morphologies store, erase, or invert directional memory under steady shear remains largely unexplored.}

\SK{In this work, we answer these questions using simulations of two-dimensional polydisperse glasses under uniform, monotonic shear. Recent work has shown that microstructural state variables can quantitatively organize macroscopic flow and memory in disordered materials \cite{galloway2022}. Motivated by this idea, we ask a complementary question under uniform, monotonic shear: can the sign and strength of directional memory (Bauschinger vs inverse Bauschinger) be predicted from preparation ($T_p$), driving ($\dot{\gamma}$), and deformation history ($\gamma_N$), and can its microscopic origin be tied to a distinct yielding mode? By scanning $T_p$ and $\dot{\gamma}$ we identify a critical history amplitude $\gamma_{N,\mathrm{crit}}(T_p,\dot{\gamma})$ that separates regimes with memory inversion from those exhibiting only conventional (classical) Bauschinger response, and we construct a phase diagram in the $(\gamma_{Nc}, T_p, \dot{\gamma})$ space.}

\SK{To connect memory to yielding, we adopt a shear-band-network viewpoint based on the spatial organization of plastic activity (e.g., from $D^2_{\min}$ fields) and track how this organization evolves with strain. We find that memory inversion correlates with network-like shear bands and rapid plastic healing, whereas conventional Bauschinger memory is associated with persistent localization and cumulative damage. This provides a microscopic mechanism for the switch in memory type and links it to the brittle-ductile character of yielding at finite strain rates. The appearance of shear-band networks also motivates a comparison with active-doped glasses, in which similar morphologies have been reported at lower rates and are argued to be analogous to passive, well-annealed glasses subjected to higher strain rates \cite{rpriya2025}; this analogy is a key motivation for our work.}

\SK{The article is organized as follows: we first describe the model, sample preparation and shear protocol. We then present the results and discussion.} 

\section{Models and Methods \protect\\}

\noindent{\bf \large Model:}
\SK{We perform simulations in a two-dimensional polydisperse model introduced in Ref.~\cite{PhysRevX.7.021039}. The particle diameters are drawn from the distribution $P(\sigma) = A/\sigma^3$, where $A$ is a normalization constant, and the mean diameter is $\bar\sigma = 1$, which sets the unit of length. The distribution is truncated between a maximum and minimum diameter, $\sigma_{\max} = 1.61 \bar \sigma$ and $\sigma_{\min}= 0.725 \bar \sigma$, respectively. The energy scale is set by $V_0 =1$, and the temperature by $k_B = 1$. The interaction between two different types of particles follows a non-additive rule for interaction and is defined as $\sigma_{ij} = \frac{\sigma_i + \sigma_j}{2} (1-0.2 |\sigma_i - \sigma_j| )$ to stablize against crytallization or demixing.}

\SK{The interaction between two particles separated by a distance $r_{ij}$ is given by a purely repulsive soft potential,
\begin{eqnarray}
V(r)=
\begin{cases}
r^{-12} + C_0 + C_2 r^2 + C_4 r^4, & r < r_c, \\
0, & r \geq r_c,
\end{cases}
\label{Potential}
\end{eqnarray}
where $r = r_{ij}/\sigma_{ij}$ and $C_0$, $C_2$, and $C_4$ are coefficients chosen to make the potential, force, and Hessian smooth at the cutoff $r_c$, and $r_c = 1.25$ is the interaction cutoff (in units of $\bar\sigma$).}

\vskip +0.1in
\noindent{\bf \large Sample preparation:}
\SK{In this work, we focus on glasses prepared across a wide range of parent temperatures and show that the nature of directional memory depends sensitively on the parent temperature. For this, the polydisperse system is annealed at different fixed parent temperatures $T_p$ ($0.026, 0.035, 0.05, 0.07, 0.10, 0.12, 0.15, 0.20$) via hybrid molecular dynamics (MD) with swap Monte Carlo (SMC) \cite{Berthier_2019, 10.1063/5.0086626}. The MD part is performed using a Nose-Hoover thermostat (chain length of 3) and the Verlet algorithm \cite{10.1093/oso/9780198803195.001.0001}. After every $25$ MD steps, $N$ swap moves are attempted, which are accepted or rejected using the Metropolis algorithm. A time step of $0.005$ is used under periodic boundary conditions. Unless otherwise stated, the system size is $N = 64{,}000$. The density of the system is fixed at $1.0$ in a square box of side length $L \approx 253$. For each $T_p$, we have performed averaging over $16$ independent ensembles, except at $T_p=0.026$, where we have used $5$.}

\begin{figure}[!htbp]
\includegraphics[width=0.49\textwidth]{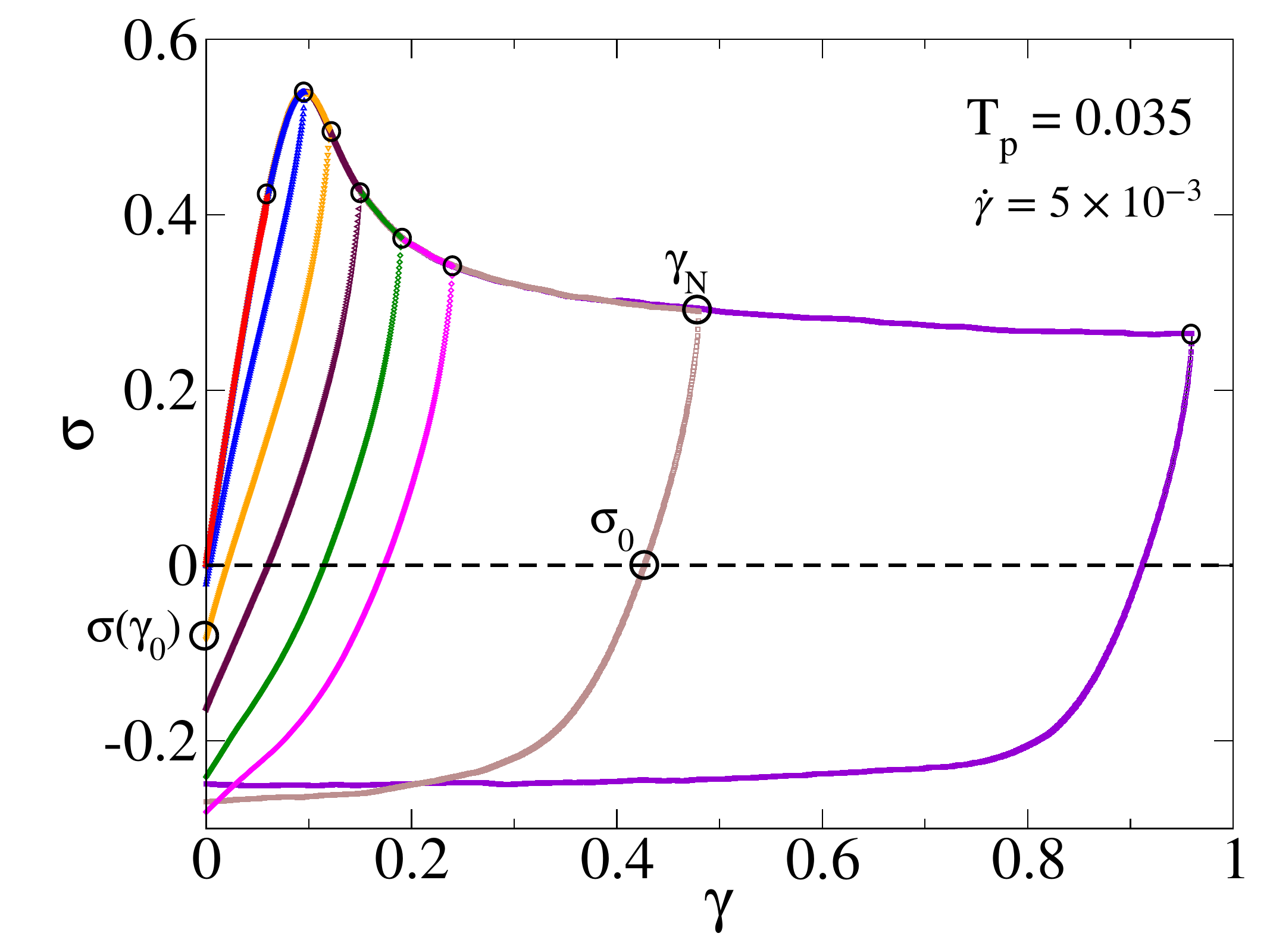}
\caption{This figure depicts the shear-reversal protocol, where a sample prepared at $T_p=0.035$ with $N=64{,}000$ is first sheared in the forward direction to different $\gamma_N$ values (denoted by circles) at a strain rate of $5\times10^{-3}$. The sample is then sheared in the reverse direction until it reaches a zero-stress configuration, labeled $\sigma_0$, and upon further shear, it reaches zero strain at $\gamma_0$, where the stress $\sigma(\gamma_0)$ is, in general, non-zero once the system has yielded.}
\label{Schematic}
\end{figure}
\begin{figure*}[htbp]
\includegraphics[width=0.95\textwidth]{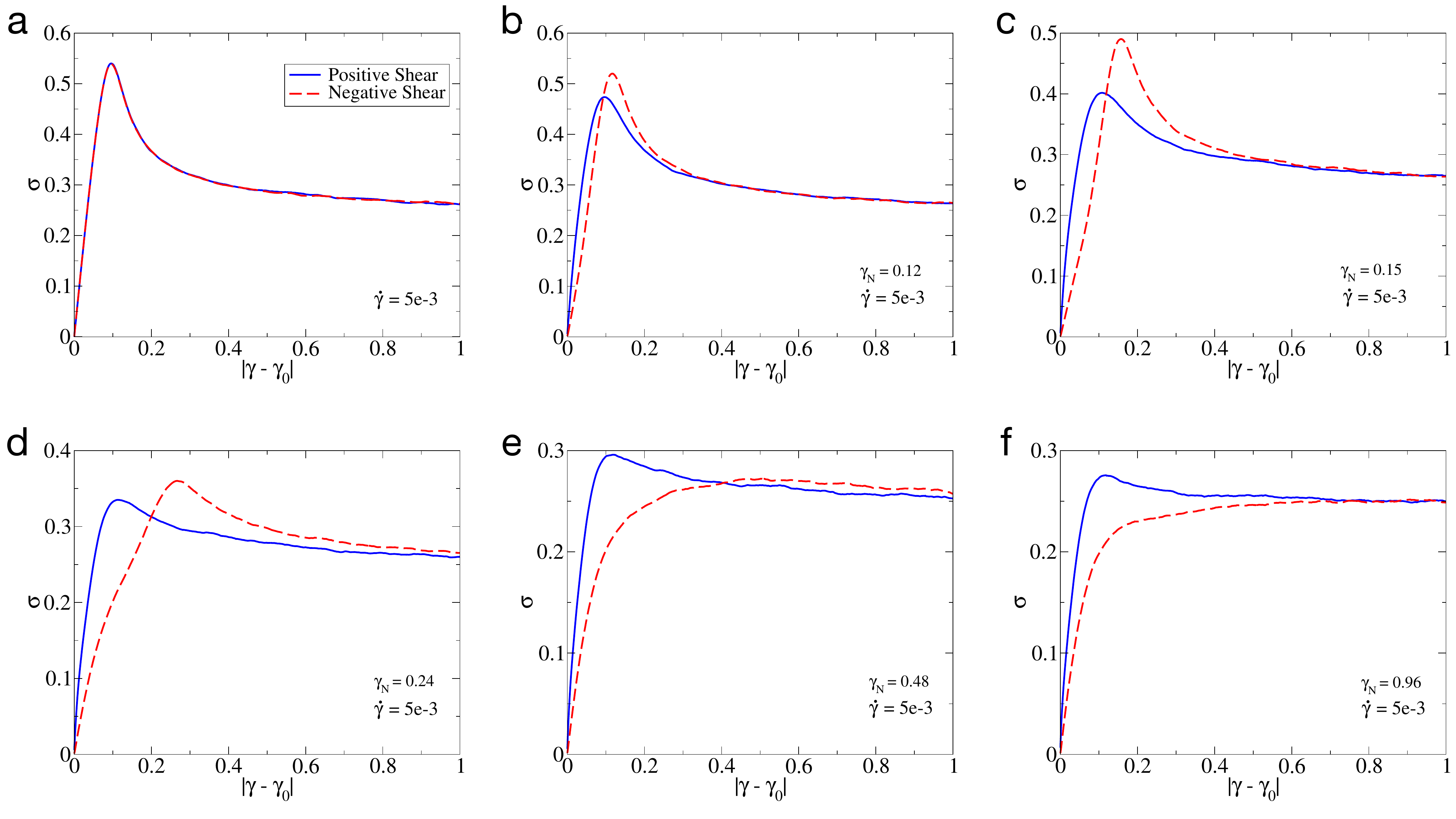}
\caption{{\bf Transition from inverse (IBE) to Bauschinger effect (BE).} Panel (a) shows a symmetric stress-strain response for a freshly prepared, well-annealed glass sample ($T_p=0.035$) sheared at a strain rate $\dot{\gamma}=5\times10^{-3}$. Panels (b-f) show the evolution of directional memory with increasing strain history. Panels (b) and (c) show pronounced yielding in the reverse direction at $\gamma_N=0.12$ and $\gamma_N=0.15$, respectively. As $\gamma_N$ approaches $0.24$ in (d), the difference between reverse and forward responses decreases. With a further increase in strain history to $\gamma_N=0.48$, the reverse yield stress has fallen below the forward yield stress, and panel (f) at $\gamma_N=0.96$ shows this more prominently, corresponding to the classical Bauschinger effect.}
\label{WellAnnealed}
\end{figure*}

\vskip +0.1in
\noindent{\bf \large Shear simulations and Shear Band Analysis:}
\SK{To shear the two-dimensional system, each prepared sample is first quenched to a low temperature $T = 0.001$. It is then subjected to different finite shear rates $\dot{\gamma}$ using the SLLOD equations of motion with a Gaussian thermostat \cite{10.1063/1.479358,10.1063/1.1858861,10.1093/oso/9780198803195.001.0001} under Lees-Edwards boundary conditions. 
Non-affine displacements were analyzed using the quantity $D^2_{\text{min}}$, which measures local deviations from an affine deformation field \cite{PhysRevE.57.7192}. This allows us to capture plastic rearrangements in amorphous solids under shear. $D^2_{\text{min}}$ is computed from the local strain tensor and quantifies the non-affine motion of each particle within a cutoff range. We choose this range as $2.5$ and use it to identify the neighbors of each particle. $D^2_{\text{min}}$ values at any non-zero strain are calculated with respect to the zero-strain configuration. For visualization, these values are rescaled to lie between $0.0$ and $2.0$: values near $0.0$ indicate regions with minimal motion or rearrangement, while values near $2.0$ highlight the most strongly rearranging regions, typically inside the shear band.}

\vskip +0.1in
\noindent{\bf \large Protocol to study unidirectional memory:}
\SK{A freshly prepared system is subjected to shear deformation in both directions up to a maximum strain value $\gamma_N$ (denoted by circles in Fig.~\ref{Schematic}) and is then brought back to a zero-stress state by reversing the shear direction (see Fig.~\ref{Schematic} for reference). This zero-stress state, labeled $\sigma_0$ in Fig.~\ref{Schematic}, generally occurs at a non-zero strain once the system has yielded. Upon further shear, the stress continues to decrease and becomes negative as zero strain ($\gamma_0$) is approached, labeled $\sigma(\gamma_0)$. We study the role of deformation history on $\sigma_0$ and $\sigma(\gamma_0)$ by varying $\gamma_N$ (see Supplementary Fig.~1 for details). The values of $\gamma_N$ are chosen according to $\gamma_N = \gamma_Y \times 10^{\pm 0.1 i}$, where $i$ is an integer in the range $[0,n]$ and $0.0 < \gamma_N < 1.0$, so as to sample more points near the yield stress/strain. Here, $\gamma_Y$ denotes the yield strain corresponding to the applied strain rate $\dot{\gamma}$. The sheared system at the zero-stress state then serves as a new initial configuration, which we subsequently shear again in the positive and negative directions to study the resulting asymmetries as a function of deformation history, $T_p$, and shear rate.}

\section{Bauschinger and inverse Bauschinger Effect}
Directional memory and its type are studied by evaluating the stress-strain response of a previously sheared sample. To do so, we follow the shear-reversal protocol described in the Methods section to set the deformation history $\gamma_N$ and obtain the corresponding zero-stress state, $(\gamma(\sigma_0),\sigma_0)$ (Fig.~\ref{Schematic}). Different strain histories return the system to zero stress at different non-zero strain values, which increase with $\gamma_N$. The system returns to zero stress at zero strain only in the elastic regime, where both fully recover. A detailed description of shear reversal in the two yielding modes (well-annealed and poorly annealed) and at different strain rates is provided in Supplementary Fig.~1. The sample is then sheared again from this zero-stress configuration ($\sigma=0$) in the same (positive) and reverse (negative) directions to evaluate the memory encoded by the prior shear.

\SK{We now highlight the stark contrast between the Bauschinger Effect (BE) and Inverse Bauschinger Effect (IBE) by comparing the stress-strain response for well-annealed and poorly annealed samples at $\dot{\gamma}=5\times10^{-3}$. Figures~\ref{WellAnnealed} and \ref{PoorlyAnnealed} show the contrasting Bauschinger and inverse Bauschinger effects, and they also show the impact of the deformation histories in both cases.
\begin{figure*}[htbp]
\includegraphics[width=0.95\textwidth]{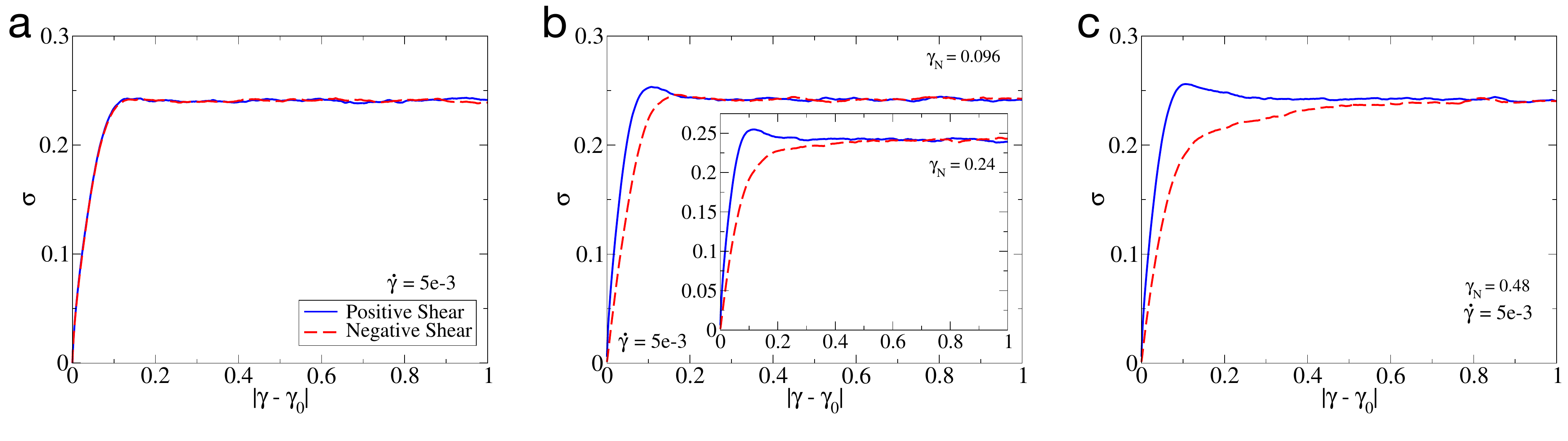}
\caption{{\bf Bauschinger effect.} Panel (a) shows a symmetric stress-strain response for a freshly prepared glass sample ($T_p=0.20$) sheared at a strain rate $\dot{\gamma}=5\times10^{-3}$. Panels (b-c) show the evolution of directional memory with increasing strain history for $\gamma_N=0.096$, $\gamma_N=0.24$ (inset of (b)), and $\gamma_N=0.48$. In all cases, the response is asymmetric, with the reverse yield stress always smaller than the forward yield stress (i.e., the direction of the previous shear), corresponding to the classical Bauschinger effect.}
\label{PoorlyAnnealed}
\end{figure*}
For a well-annealed sample ($T_p=0.035$), a freshly prepared system shows a symmetric stress response upon shearing in the positive ($+x$) and negative ($-x$) directions, as shown in Fig.~\ref{WellAnnealed}(a). When the same system is first sheared in the positive/forward direction to a maximum strain $\gamma_N$, brought back to a zero-stress state at strain $\gamma(\sigma_0)$, and then re-sheared from this configuration in both forward and reverse directions, an asymmetry in the stress-strain response emerges, and its nature changes with increasing deformation history (Fig.~\ref{WellAnnealed}(b-f)).}

\SK{Figure~\ref{WellAnnealed}(b) shows that at $\gamma_N=0.12$ the negative direction exhibits a higher yield stress (and a larger yield strain) than the positive direction, indicating inverse Bauschinger response. This difference increases further at $\gamma_N=0.15$ (Fig.~\ref{WellAnnealed}(c)).} Upon increasing the deformation history further, e.g., near $\gamma_N \approx 0.24$, the negative-direction yield stress remains larger than the positive-direction yield stress, but the difference between them decreases. \SK{The two yield stresses become equal at the transition strain $\gamma_N=\gamma_{Nc}$ (as identified in the Fig.~\ref{degree_of_Annealing} (a) and (b)), and for larger deformation histories the response crosses over to the conventional Bauschinger effect, e.g., at $\gamma_N=0.48$ and $0.96$.}

\SK{For a poorly annealed sample ($T_p=0.2$) at the same strain rate $\dot{\gamma}=5\times10^{-3}$, the evolution is qualitatively different. A freshly prepared system again shows a symmetric response in opposite shear directions, as shown in Fig.~\ref{PoorlyAnnealed}(a), and an asymmetry appears after imposing a prior forward deformation. However, over the entire range of deformation histories explored, the asymmetry corresponds only to the conventional Bauschinger effect: the yield stress in the direction opposite to the prior shear remains lower than in the forward direction, and the separation between the two curves grows as $\gamma_N$ increases from $0.096$ to $0.96$, reflecting the presence of a pronounced Bauschinger effect. This contrasts with the well-annealed case, where the system exhibits an IBE-to-BE crossover with increasing deformation history, reflecting the presence of memory inversion in well-annealed samples as opposed to poorly annealed ones.}

\section{Effect of thermal history and shear rate on IBE to BE Transition} 
\begin{figure*}[!htp]
\includegraphics[width=0.99\textwidth]{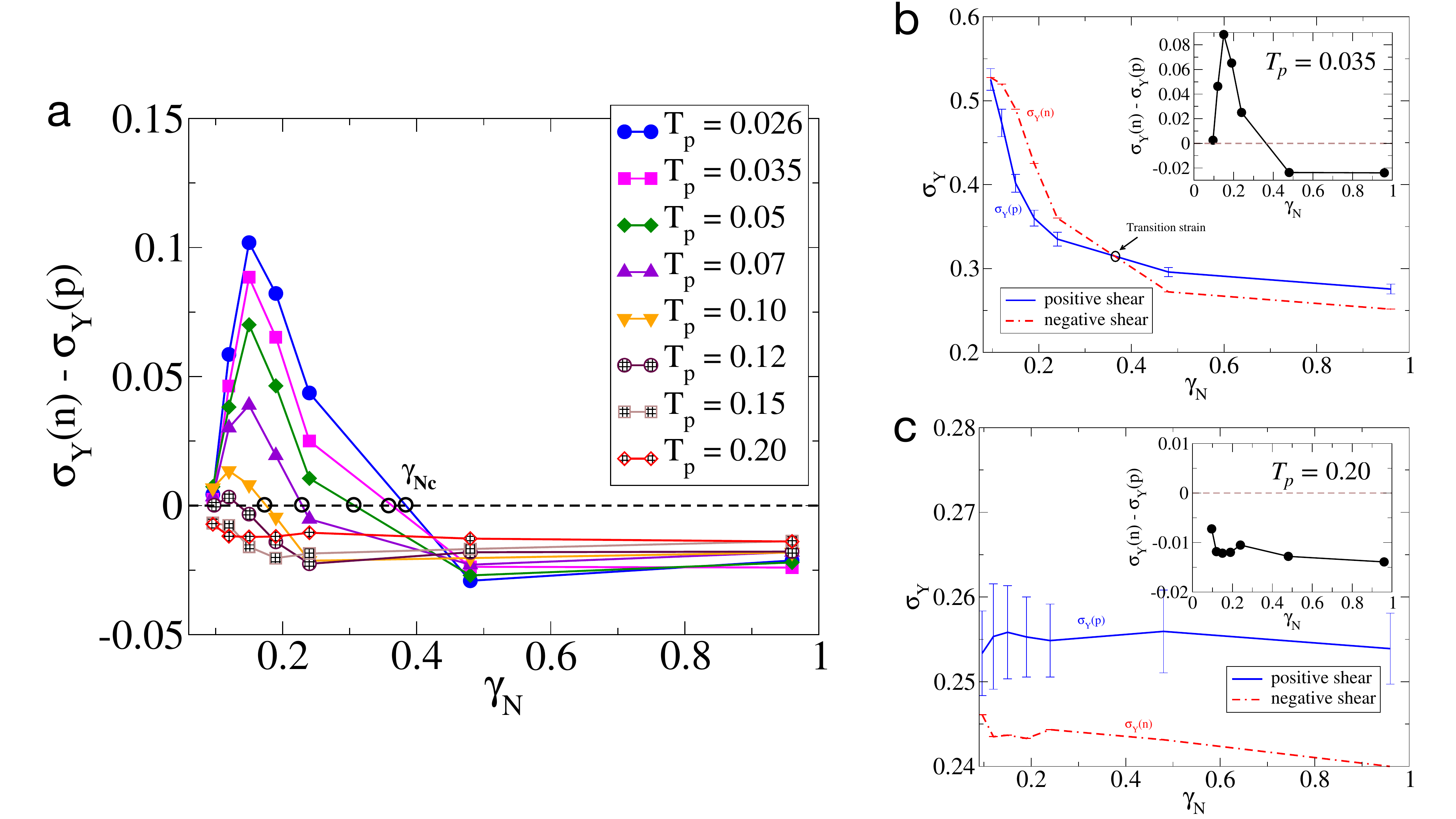}
\caption{{\bf Annealing dependence of directional memory.} (a) The panel shows the difference between the yield stresses measured when the system is sheared in the previously sheared direction to a maximum strain $\gamma_N$, $\sigma_Y(p)$, and in the opposite direction, $\sigma_Y(n)$. We observe a transition from IBE ($\sigma_Y(n)-\sigma_Y(p)>0$) to BE ($\sigma_Y(n)-\sigma_Y(p)<0$) as $\gamma_N$ increases beyond $\gamma_{Nc}$ at lower parent temperatures, $T_p \leq 0.10$. As the parent temperature increases, the peak height decreases and the IBE regime with $\sigma_Y(n)-\sigma_Y(p)>0$ shrinks, eventually giving way to an always BE regime with $\sigma_Y(n)-\sigma_Y(p)<0$ around $T_p \approx 0.12$. Panels (b) and (c) illustrate the evolution of yield stresses as a function of deformation history $\gamma_N$ at two parent temperatures, $T_p=0.035$ and $T_p=0.20$, measured in the positive (blue) and negative (red) shear directions; the insets show their difference, $\sigma_Y(n)-\sigma_Y(p)$, as in panel (a). The low-$T_p$ sample exhibits a clear inverse Bauschinger regime with a well-defined transition strain $\gamma_{Nc}$, whereas this regime is absent for the poorly annealed glass at $T_p=0.20$.}
\label{degree_of_Annealing}
\end{figure*}
\noindent{\bf \large Annealing Effect:} 
To study the effect of thermal history on memory, we apply the shear-reversal protocol described above to glasses prepared across a wide range of stabilities (annealing), controlled by the parent temperature $T_p = 0.026, 0.035, 0.05, 0.07, 0.1, 0.12, 0.15,$ and $0.2$. \SK{For both directions, the new yield stresses are denoted $\sigma_Y(p)$ for positive shear and $\sigma_Y(n)$ for negative shear, extracted as the first maximum in the stress-strain curve at yield (see Supplementary Fig.~2 for details). We define an order parameter $\Delta \sigma_{BE}$: 
\begin{equation}
\Delta \sigma_{BE} = \sigma_Y(n) - \sigma_Y(p),
\label{eq_M}
\end{equation}
as the difference in the yield stress values, which captures both the asymmetry and its sign (i.e., the type of directional memory) arising from the previous loading history (encoded memory).
}

\SK{The inverse Bauschinger effect (IBE) exists if $\sigma_Y(n) > \sigma_Y(p)$ (i.e., $\Delta \sigma_{BE}>0$), and the Bauschinger effect (BE) exists when $\sigma_Y(n) < \sigma_Y(p)$ (i.e., $\Delta \sigma_{BE}<0$). $\Delta \sigma_{BE}$ changes sign in some cases (IBE to BE) and not in others (only BE) as the strain history increases at a fixed strain rate $\dot{\gamma}=5\times 10^{-3}$. The yield-stress difference varies with deformation history $\gamma_N$ and parent temperature $T_p$. For well-annealed (low-$T_p$) samples, $\Delta \sigma_{BE}$ first increases with $\gamma_N$, reaches a positive maximum, and then decreases, crossing zero and becoming negative as $\gamma_N$ increases. Thus, at fixed $T_p$, increasing $\gamma_N$ drives a transition from IBE to BE. The peak height systematically decreases with increasing $T_p$, while the regime where IBE is observed simultaneously shrinks, as shown in Fig.~\ref{degree_of_Annealing}(a). For $T_p \gtrsim 0.1$, $\Delta \sigma_{BE}$ stays negative for all $\gamma_N$ explored (no IBE regime is observed), and the memory remains BE.} This illustrates how the transition occurs as the preparation temperature increases, from well-annealed to poorly annealed samples, with the inverse Bauschinger effect serving as a directional-memory order parameter across a crossover around $T_p \approx 0.1$ at the strain rate studied here. Notably, this crossover lies close to the random critical temperature reported previously for the brittle-ductile transition (2D) in the AQS limit \cite{article}, although the precise crossover temperature is strain-rate dependent.

\SK{Figures~\ref{degree_of_Annealing}(b) and (c) show the evolution of the yield stresses for a well-annealed sample ($T_p=0.035$) and a poorly annealed sample ($T_p=0.2$) in the positive (blue curve) and negative (red curve) shear directions, with the inset showing their difference, as in (a), for clarity. Panel (b) clearly indicates a transition strain history where the curve ordering switches: the red curve goes from staying above the blue curve to decreasing below it, indicating a transition from IBE to BE. This is completely absent in (c), where the blue curve (positive direction) always stays above the red curve (negative direction).}
 
\vskip +0.1in
\noindent{\bf \large Shear-rate Effect:}
\begin{figure*}[htbp]
\includegraphics[width=0.99\textwidth]{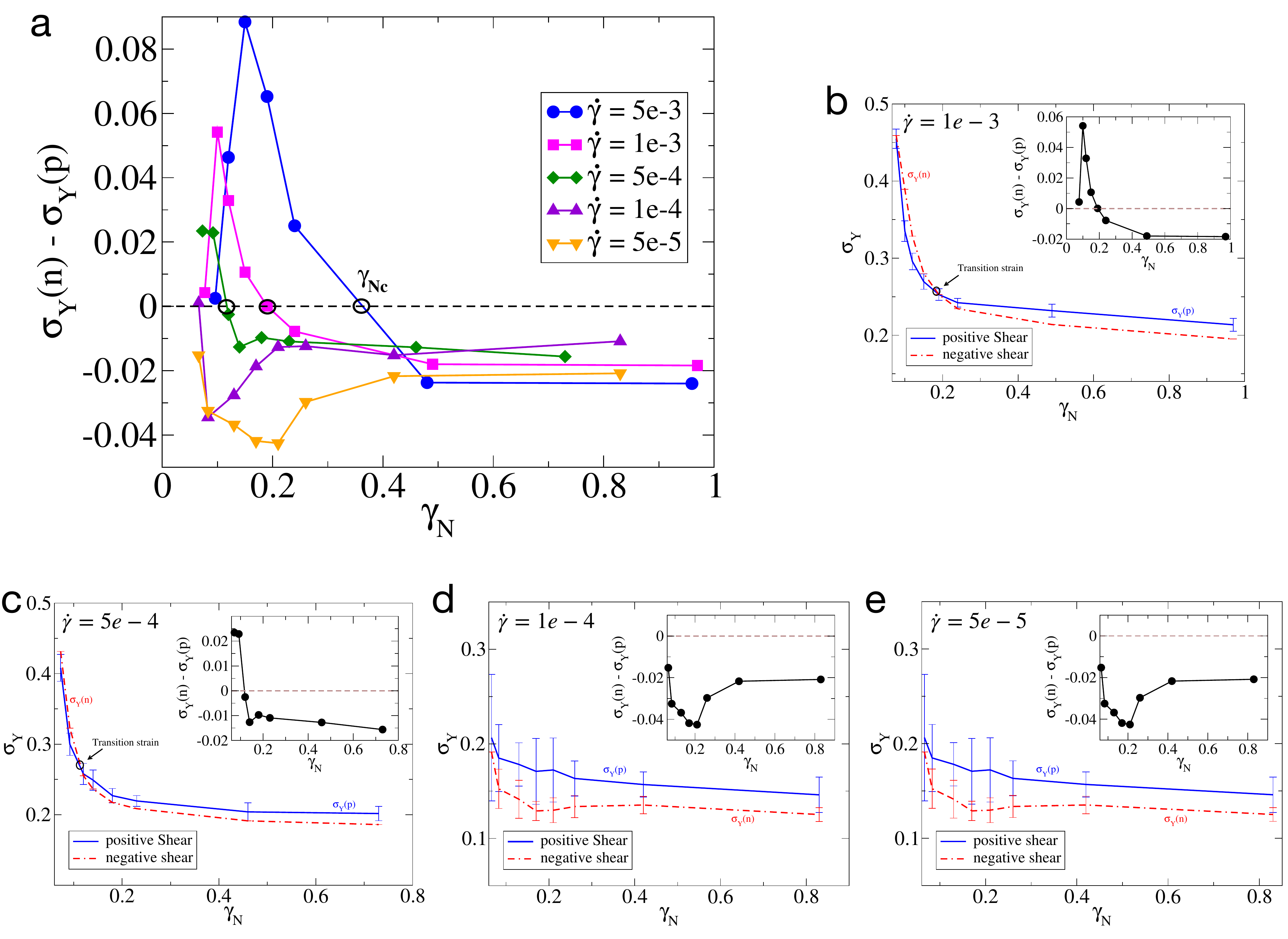}
\caption{{\bf Finite shear rate and directional memory.} (a) The panel shows the difference between the yield stresses, $\sigma_Y(n)-\sigma_Y(p)$, as a function of the strain history $\gamma_N$. As the strain rate decreases, the peak height decreases and the IBE regime with $\sigma_Y(n)-\sigma_Y(p)>0$ shrinks, reducing $\gamma_{Nc}$ (=$\gamma_{N,\mathrm{crit}}(T_p,\dot{\gamma})$) and tending to zero around $\dot{\gamma} \approx 1\times10^{-4}$. Panels (b-e) show yield stresses as a function of deformation history $\gamma_N$ at $T_p=0.035$ for decreasing strain rates, $\dot{\gamma}=1\times10^{-3}$, $5\times10^{-4}$, $1\times10^{-4}$, and $5\times10^{-5}$, measured in the positive (blue) and negative (red) shear directions; the insets show their difference ($\sigma_Y(n)-\sigma_Y(p)$) as in (a). The higher strain rates $\dot{\gamma}=5\times10^{-3}$ (see Fig.~\ref{degree_of_Annealing}(b)) and $\dot{\gamma}=1\times10^{-3}$ (panel (b)) exhibit a clear inverse Bauschinger regime with a well-defined transition, which becomes very small at $\dot{\gamma}=5\times10^{-4}$ and crosses over to the Bauschinger regime as $\gamma_N$ increases. At the lower strain rates $\dot{\gamma}=1\times10^{-4}$ and $5\times10^{-5}$, the inverse-Bauschinger regime is completely absent over the entire range of $\gamma_N$ studied.}
\label{strainrate}
\end{figure*}
%
\SK{In the previous section, we discussed the effect of annealing on directional memory at a fixed, high strain rate. Here, we use the fact that even at high strain rates, a poorly annealed sample shows only the conventional Bauschinger effect, and therefore we restrict ourselves to the well-annealed sample $T_p=0.035$ to study the strain-rate dependence of directional memory and the IBE-BE crossover (see Supplementary Fig.~3 for the corresponding stress-strain curve). We compute the same order parameter $\Delta \sigma_{BE}$, defined in Eq.~\ref{eq_M}, as the difference between the yield stresses in the two shear directions. At the faster strain rate $\dot{\gamma}=5\times 10^{-3}$, Fig.~\ref{degree_of_Annealing}(b) shows a clear IBE regime, where the yield stress in the negative direction exceeds the positive one, with a well-defined transition strain history $\gamma_{Nc}\simeq 0.36$ at which the difference goes to zero. We repeat the same analysis for other strain rates in Fig.~\ref{strainrate}(b-e), where the insets show the corresponding differences in yield stresses between the two directions, and compile the results in Fig.~\ref{strainrate}(a) for a direct comparison.}

\SK{At higher strain rates, the yield-stress curves in the two probe directions show the same qualitative behavior as in the annealing-controlled case: initially, the negative-direction yield stress becomes larger than the positive-direction one, i.e., $\sigma_Y(n)>\sigma_Y(p)$, giving a positive window corresponding to IBE. As $\gamma_N$ is increased further, this difference decreases, crosses zero at a well-defined transition strain $\gamma_{Nc}$, and the system re-enters the conventional Bauschinger regime with $\sigma_Y(n)<\sigma_Y(p)$. This IBE$\to$BE crossover is clearly visible at $\dot{\gamma}=5\times10^{-3}$ (Fig.~\ref{degree_of_Annealing}(b)) and persists at $\dot{\gamma}=1\times10^{-3}$ (Fig.~\ref{strainrate}(b)), although both the peak value and the extent of the IBE window in $\gamma_N$ are reduced. At $\dot{\gamma}=5\times10^{-4}$ (Fig.~\ref{strainrate}(c)), the IBE regime is much smaller, with the transition occurring at a very small $\gamma_{Nc}$. Upon decreasing the strain rate further, the IBE window collapses: for $\dot{\gamma}=1\times10^{-4}$ and $5\times10^{-5}$ (Fig.~\ref{strainrate}(d,e)), the negative-direction yield stress never exceeds the positive one for any $\gamma_N$ explored, so $\sigma_Y(n)-\sigma_Y(p)$ remains negative and only the conventional Bauschinger effect is observed. This trend is summarized in Fig.~\ref{strainrate}(a): lowering $\dot{\gamma}$ systematically suppresses the positive peak in $\sigma_Y(n)-\sigma_Y(p)$, shrinks the range of $\gamma_N$ where it is positive, and shifts the transition strain $\gamma_{Nc}$ to smaller values, until $\gamma_{Nc}\to 0$ around $\dot{\gamma}\approx 10^{-4}$.}

\vskip +0.1in
\noindent{\bf \large State Diagram of IBE-BE Crossover in the $(T_p,\dot{\gamma})$ Plane:}
\SK{Having established the separate roles of preparation $T_p$ and shear rate $\dot{\gamma}$, we now present a unified classification of inverse and conventional Bauschinger response in the $(T_p,\dot{\gamma})$ plane. For each $(T_p,\dot{\gamma})$, we determine the critical deformation history $\gamma_{N,\mathrm{crit}}(T_p,\dot{\gamma})$=$\gamma_{Nc}$ at which the directional-memory order parameter $\Delta \sigma_{BE}=\sigma_Y(n)-\sigma_Y(p)$ changes sign, i.e., where IBE and BE yield the same yield stress ($\Delta \sigma_{BE}=0$). The value of $\gamma_{Nc}$ is obtained by interpolating the $\Delta \sigma_{BE}(\gamma_N)$ curves discussed above (see also Fig.~\ref{strainrate}(a)). When a clear IBE$\to$BE transition exists, we obtain a non-zero $\gamma_{Nc}\neq 0$; when no sign change is observed within the explored $\gamma_N$ range, we set $\gamma_{Nc}=0$. Using these values, we construct the heat map shown in Fig.~\ref{Phasediag}.
\begin{figure}[htbp]
\includegraphics[width=0.49\textwidth]{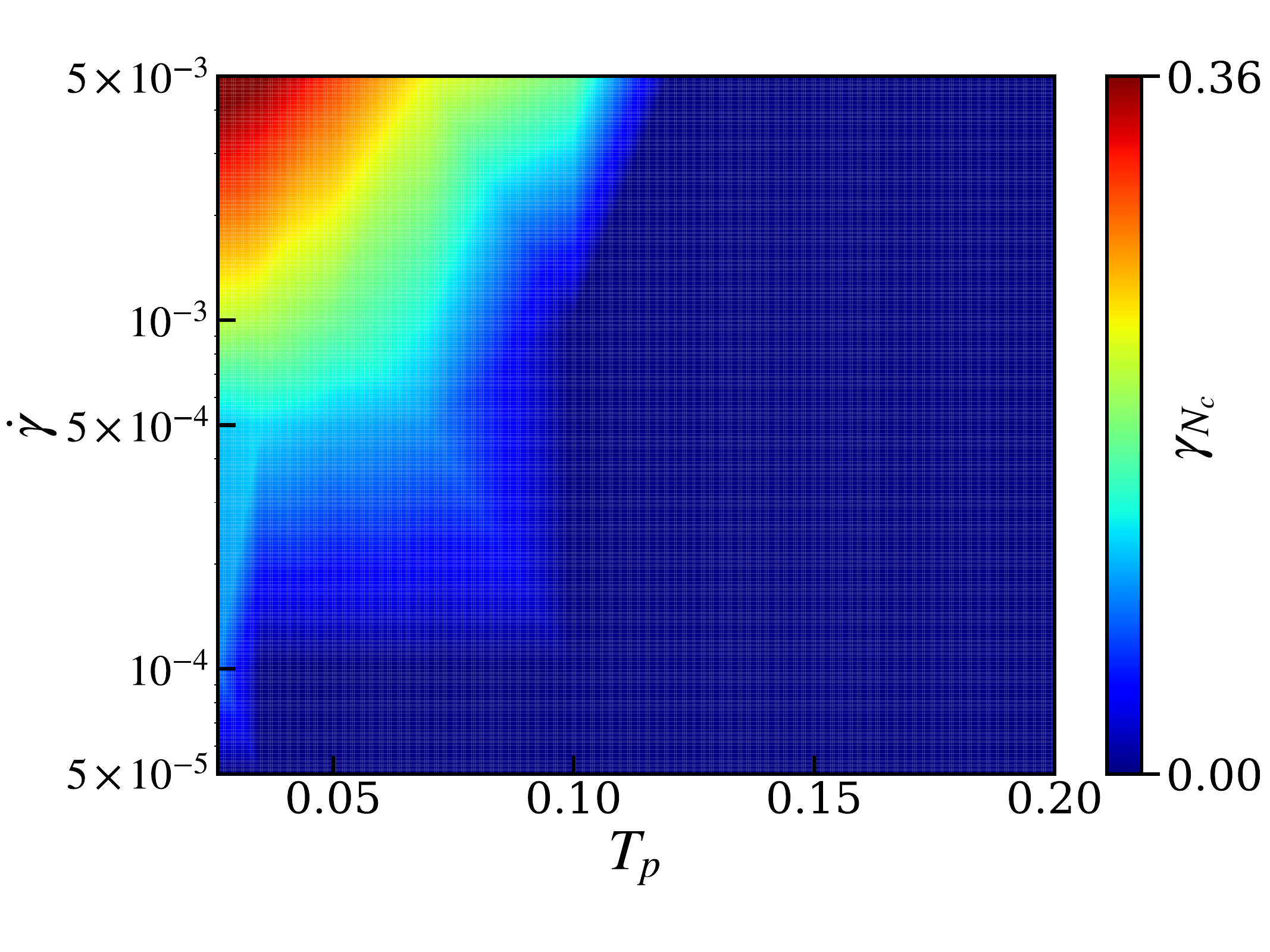}
\caption{{\bf Critical deformation history for unidirectional memory type.} Heat map of the IBE-to-BE transition history $\gamma_{Nc}$ ($\gamma_{Nc}$ = $\gamma_{N,\mathrm{crit}}(T_p,\dot{\gamma})$) in the $(T_p,\dot{\gamma})$ plane, obtained from the sign change of the directional-memory order parameter $\Delta \sigma_{BE}=\sigma_Y(n)-\sigma_Y(p)$. For a given $(T_p,\dot{\gamma})$, an inverse Bauschinger response is observed only for a non-zero critical deformation history $\gamma_{Nc}$ (colored region), whereas $\gamma_{Nc}=0$ (blue) indicates that only the conventional Bauschinger effect is observed and no IBE occurs.}
\label{Phasediag}
\end{figure}
This two-dimensional map uses $T_p$ (x-axis) and $\dot{\gamma}$ (y-axis), with color indicating the critical deformation history magnitude $\gamma_{Nc}$ from $0.0$ to $0.36$. Decreasing $T_p$ corresponds to deeper annealing and, in our system, is accompanied by a ductile-to-brittle change in the yielding response. The blue region ($\gamma_{Nc}=0$) corresponds to conditions where no inverse Bauschinger regime is observed, consistent with the suppression of IBE at higher $T_p$, lower $\dot{\gamma}$, or both.}

\SK{In the remaining region, $\gamma_{Nc}$ increases as $T_p$ decreases and/or $\dot{\gamma}$ increases, indicating that well-annealed glasses driven at faster rates sustain an extended IBE window before crossing over to conventional Bauschinger response at larger deformation histories. Within our explored parameter window, the maximum value $\gamma_{Nc}\simeq 0.36$ occurs at $T_p=0.026$} (ultrastable regime) \SK{and $\dot{\gamma}=5\times10^{-3}$. This systematic organization of $\gamma_{Nc}(T_p,\dot{\gamma})$ motivates the microscopic analysis in the next section, where we connect the emergence of IBE to the yielding mode and the associated plastic rearrangements.}

\section{Microscopic origin: plastic healing vs cumulative damage}
\SK{To understand the reason behind the contrasting effects in the two regimes, we turn to analysing the detailed structure of shear bands, since this is the key feature that differs between the yielding modes. Brittle-like yielding results in a well-defined system-spanning shear band or a shear-band network at faster strain rates, which is not true for ductile-like yielding in poorly annealed glass. Poorly annealed glass exhibits only weak, diffuse plastic activity at both slower and faster strain rates. We therefore quantify shear bands as a structural transition metric using $D^2_{\rm min}$ fields (see Methods section). It has been observed earlier in the active-particle-doped case that healing of a shear-band network occurs upon reversing the direction, which leads to delayed yielding and an increased yield stress \cite{rpriya_IBE}. Here, we observe that a similar mechanism is responsible for the distinction between the two yielding modes.}
\begin{figure*}[htpb]
\includegraphics[width=0.99\textwidth]{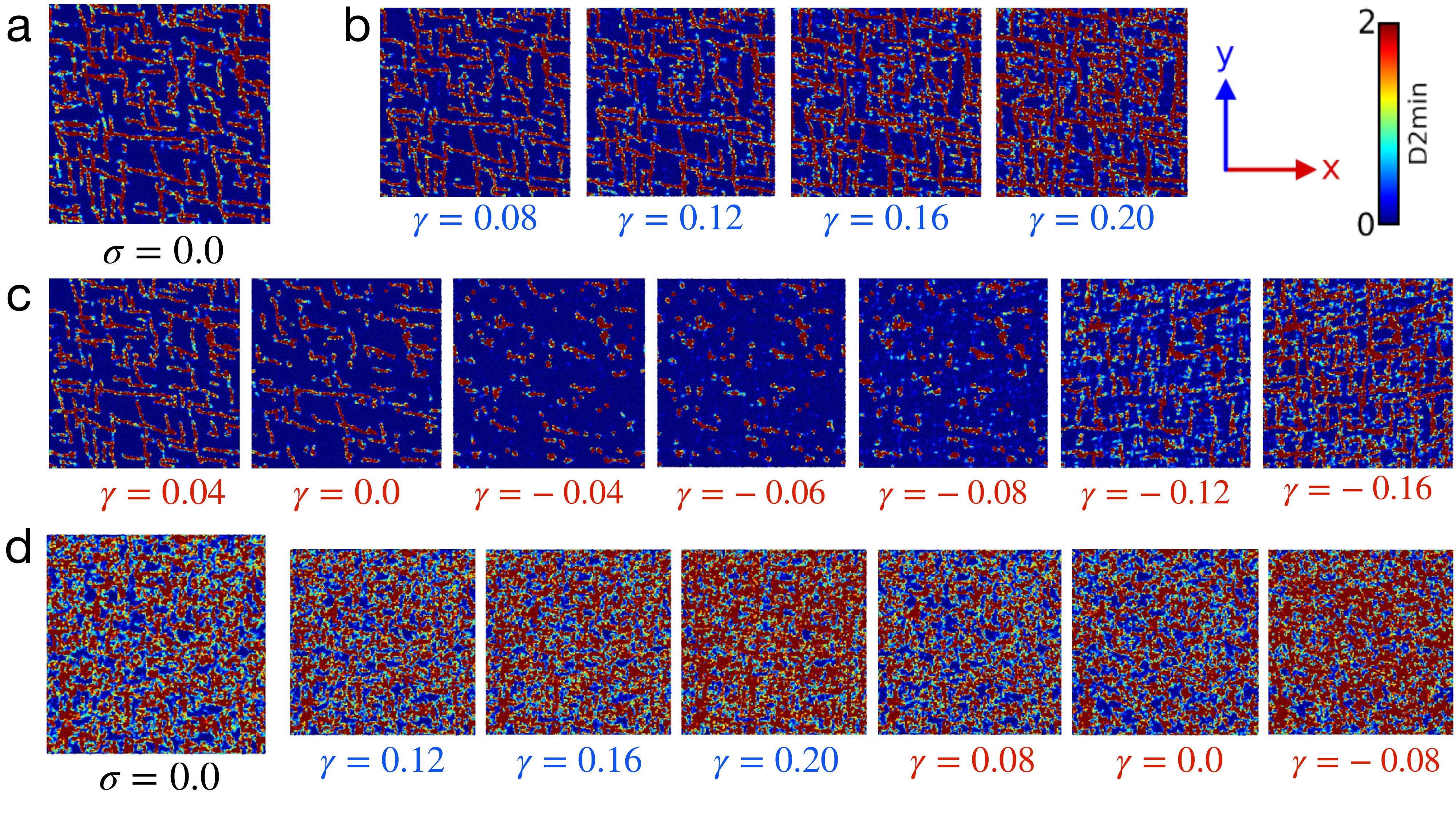}
\caption{{\bf Plastic healing underlies the inverse Bauschinger effect.} Panels (a-c) refer to a well-annealed glass ($T_p=0.035$), whereas panel (d) refers to a poorly annealed glass ($T_p=0.2$) for a deformation history of $\gamma_N=0.15$. (a) shows the non-affine field at the zero-stress state ($\gamma(\sigma_0)=0.062$, $\sigma=0.0$) for the well-annealed sample under shear at $\dot{\gamma}=5\times10^{-3}$. (b) shows shear-band evolution in the same direction as the previous shear (blue label). (c) shows shear-band evolution in the opposite direction to the original shear (red label). It is evident that, in (b), the shear band propagates from the soft region highlighted in (a), whereas in (c), the plasticity heals up to a point and then yields again, forming a shear-band network around the old soft zones near $\gamma=-0.08$. This healing is absent in both the forward (blue label) and reverse directions (red labels) for the poorly annealed glass in (d), where there is no well-defined shear-band network and the plastic activity is more homogeneously diffused. The strain values are measured relative to the strain at the zero-stress state.}
\label{Healing}
\end{figure*}

\SK{Figure~\ref{Healing}(a-c) is for a well-annealed sample prepared at $T_p=0.035$ and strain rate $\dot{\gamma}=5\times10^{-3}$, for deformation history $\gamma_N=0.15$. Panel (a) shows the zero-stress state $\sigma_0$ at strain $\gamma(\sigma_0)\approx 0.062$. Panel (b) shows the evolution of non-affine displacement in the same direction when shear is continued (blue labels): it highlights the shear-band network and its evolution as strain progresses. Panel (c) shows the response after reversing the shear direction (red labels). As reverse strain progresses, the shear-band network diminishes, as seen from the vanishing red-colored regions, until $\gamma \approx -0.06$, where only non-connected soft regions remain. Around $\gamma \approx -0.08$, new soft regions emerge and the system yields. With further increase in the magnitude of reverse strain, these regions connect to form a new interconnected shear-band network.} Thus, the observed inverse Bauschinger effect arises from a network-like shear-band structure that undergoes healing during shear reversal, a phenomenon absent when shearing persists in the original direction. \SK{Faster strain rates yield gradually via a shear-band network (which can partially heal upon reversal), whereas at slower strain rates, yielding localizes abruptly into a single system-spanning band that becomes imprinted and is difficult to heal.}

\SK{At low strain $\gamma$ and at high strain rates, plastic activity organizes into a sparse, still-developing network of shear-band segments rather than a deep, fully percolated system-spanning band. With further increase in strain, this fragmented network gradually connects and eventually spans the system. Once a spanning network forms, the associated anisotropy is imprinted, plastic healing upon reversal becomes ineffective, and the response crosses over to the conventional Bauschinger effect. Therefore, the most favorable window for observing the inverse Bauschinger effect is near yield in the transient regime (can be seen in Fig.~\ref{degree_of_Annealing}), where the system has not yet reached steady flow, and the shear-band pattern remains a non-percolated, reconfigurable network. The same shear-band morphology, its evolution, and the underlying plastic healing are also observed for the smaller system size $N=10{,}000$, provided the glass is well-annealed or ultrastable (see Supplementary Fig.~4).}

\SK{In contrast, at slow strain rates in well-annealed glasses, yielding occurs abruptly through a single system-spanning shear band that forms at low yield strain and concentrates most subsequent plastic activity, leaving the rest of the system mechanically locked. Upon reversal, deformation predominantly reactivates this same softened band, where rearrangements are easier, while the surrounding material remains comparatively rigid. This strong localization limits spatially distributed rearrangements across the system that could erase the directional bias and therefore promotes the conventional Bauschinger effect \cite{rpriya_IBE}.} 

The inverse Bauschinger effect has also been reported previously in active-doped systems \cite{rpriya_IBE}. \SK{However, our results reveal that the presence of the inverse Bauschinger effect in amorphous solids is governed primarily by shear-band morphology and the associated failure mode, rather than by activity itself. Consistent with this, the IBE is observed at high strain rates even in passive systems, indicating that active doping is not a necessary prerequisite. Activity can nevertheless play a compensatory role, where active forces effectively offset strain-rate effects at low $\tau_p$, thereby promoting more gradual failure and multiple shear-band networks at strain rates where the passive system would otherwise show abrupt localization \cite{rpriya_IBE, rpriya2025}. Thus, active doping can extend the region of the phase diagram (Fig.~\ref{Phasediag}) over which the IBE is observed.}

\SK{The same analysis in poorly annealed glass (Fig.~\ref{Healing} (d)) shows no well-defined shear bands or network of bands; instead, they are diffuse and spread homogeneously throughout the system, and even upon increasing or decreasing strain rates, they do not show much difference in plastic rearrangements \cite{PhysRevMaterials.4.025603}. We show the zero-stress state in Fig.~\ref{Healing}(d) and the respective evolution: positive-direction strain values are indicated by blue labels and negative-direction strain by red labels. In both directions, we do not observe a clear directional signature of healing: the non-affine regions (higher $D^2_{\rm min}$) remain diffuse and homogeneous, without the formation/erasure of an interconnected network upon reversal.}

\section{Discussion}
\SK{We have studied unidirectional memory and its relation to the stability of the glass, controlled by annealing (preparation temperature) and finite shear rate. We provide a unified framework linking the type of unidirectional memory (Bauschinger versus inverse Bauschinger) to the brittle-ductile character of yielding under steady shear, and show how this link is controlled by preparation and driving. We identify three control parameters-parent temperature $T_p$ (a proxy for kinetic stability), applied shear rate $\dot{\gamma}$, and deformation history $\gamma_N$ and summarize them in a phase diagram. The sign of the directional-memory order parameter $\Delta\sigma_{BE}$ acts as a state variable for the brittle-ductile yielding crossover, determined by a critical history amplitude $\gamma_{Nc}$ and associated with a transition in plastic organization from persistent localization to a networked shear-band response. We identify plastic healing as the microscopic origin of memory inversion and show that, for each $(T_p,\dot{\gamma})$, the inverse Bauschinger effect disappears beyond a critical deformation history $\gamma_{Nc}$, marking the crossover between brittle-like and ductile-like yielding at finite strain rates. This establishes directional memory as an additional order parameter for the brittle-ductile crossover and provides a new way to detect it, complementary to stress-drop statistics.} 

\SK{Bauschinger-like asymmetries, shear-band localization, and networked banding are widely observed across metallic glasses, polymer glasses, and dense soft amorphous materials. While the Bauschinger effect has already been characterized in sheared dense colloidal suspensions \cite{PhysRevE.106.034611} and in Carbopol suspensions, where the history dependence is linked to the emergence of shear banding \cite{PhysRevLett.110.018304}, the ubiquity of these signatures suggests that our predicted phenomena, plastic healing and the inverse Bauschinger response, should be equally accessible to experimental tests. Controlled pre-shear and reversal protocols provide a direct route to probe these signatures and to validate the role of directional memory in yielding at finite shear rates. More broadly, such memory suggests a route for materials design, where deformation histories tune anisotropy and localization tendencies in soft precursors before solidification \cite{divoux2024ductile}, thereby directly linking memory in amorphous solids to their yielding and failure modes at finite shear rates.}

\SK{Our results also clarify the role of shear-band structure in promoting gradual yielding and enlarging the window over which memory can be probed near the yielding transition. The emergence of a shear-band network introduces a characteristic length scale that has been argued to originate from the competition between the imposed shear rate and the spatiotemporal propagation of correlations mediated by Eshelby-like shear-transformation events \cite{PhysRevMaterials.4.025603}. More broadly, inverse Bauschinger response correlates with network-like banding and rapid plastic healing, whereas conventional Bauschinger response is associated with persistent localization and cumulative damage in well-annealed systems. In poorly annealed systems, plastic activity is mostly spatially diffuse, with no persistent or well-defined shear-band network. Consequently, local healing is overwhelmed by ongoing rearrangements arising from increased mobility and the ability of particles to redistribute stress throughout the sample, thereby suppressing the inverse Bauschinger effect.} 

\SK{Our results also indicate that shear-band morphology, rather than active doping itself, is the key control variable. Activity acts indirectly by promoting networked banding and compensating for shear-rate effects; for instance, the Inverse Bauschinger Effect (IBE) emerges at much lower strain rates in active-doped systems compared to their passive counterparts. Crucially, if a specific pair of active force and strain rate produces matched macroscopic stress and inherent-state energies, it should yield an identical memory response. This suggests that the microscopic structural state, once formed, dictates the memory type regardless of the specific combination of driving forces used to reach it.} 

\SK{Finally, our results highlight the importance of post-yield history for material longevity. The same plastic-healing mechanism that governs the crossover between BE and IBE under simple shear is expected to also operate under oscillatory loading. 
Although we present results for two-dimensional polydisperse glasses, the ingredients underlying plastic healing and directional memory, such as localization, shear-band formation, and rate-controlled networking, are generic and should extend to other interaction models and to three-dimensional systems. By demonstrating how microstructural rearrangements encode directional memory, our findings provide a framework for tailoring the response of disordered materials by controlling annealing and strain-rate parameters.}

\vskip +0.05in
\begin{acknowledgments}
We acknowledge the funding by intramural funds at TIFR Hyderabad from the Department of Atomic Energy (DAE) under Project Identification No. RTI 4007. SK would like to acknowledge Swarna Jayanti Fellowship Grant Nos. DST/SJF/PSA01/2018-19 and SB/SFJ/2019-20/05 from the Science and Engineering Research Board (SERB) and Department of Science and Technology (DST). SK also acknowledges research support from MATRICES Grant MTR/2023/000079 from SERB.
\end{acknowledgments}

\bibliography{inverseBE}

\end{document}


\title{\textbf{Supplementary Information - Inverse Bauschinger to Bauschinger Crossover under Steady Shear in Amorphous Solids}} 

\author{Rashmi Priya$^1$} 
\author{Smarajit Karmakar$^1$} 
\affiliation{ $^1$ Tata Institute of Fundamental Research, 36/P, Gopanpally Village, Serilingampally Mandal,
Ranga Reddy District, Hyderabad, India 50046}
\maketitle

\section{Non-zero Residual stress and signatures of directional memory}

\begin{figure}[htb]
\centering
\includegraphics[width=0.95\textwidth]{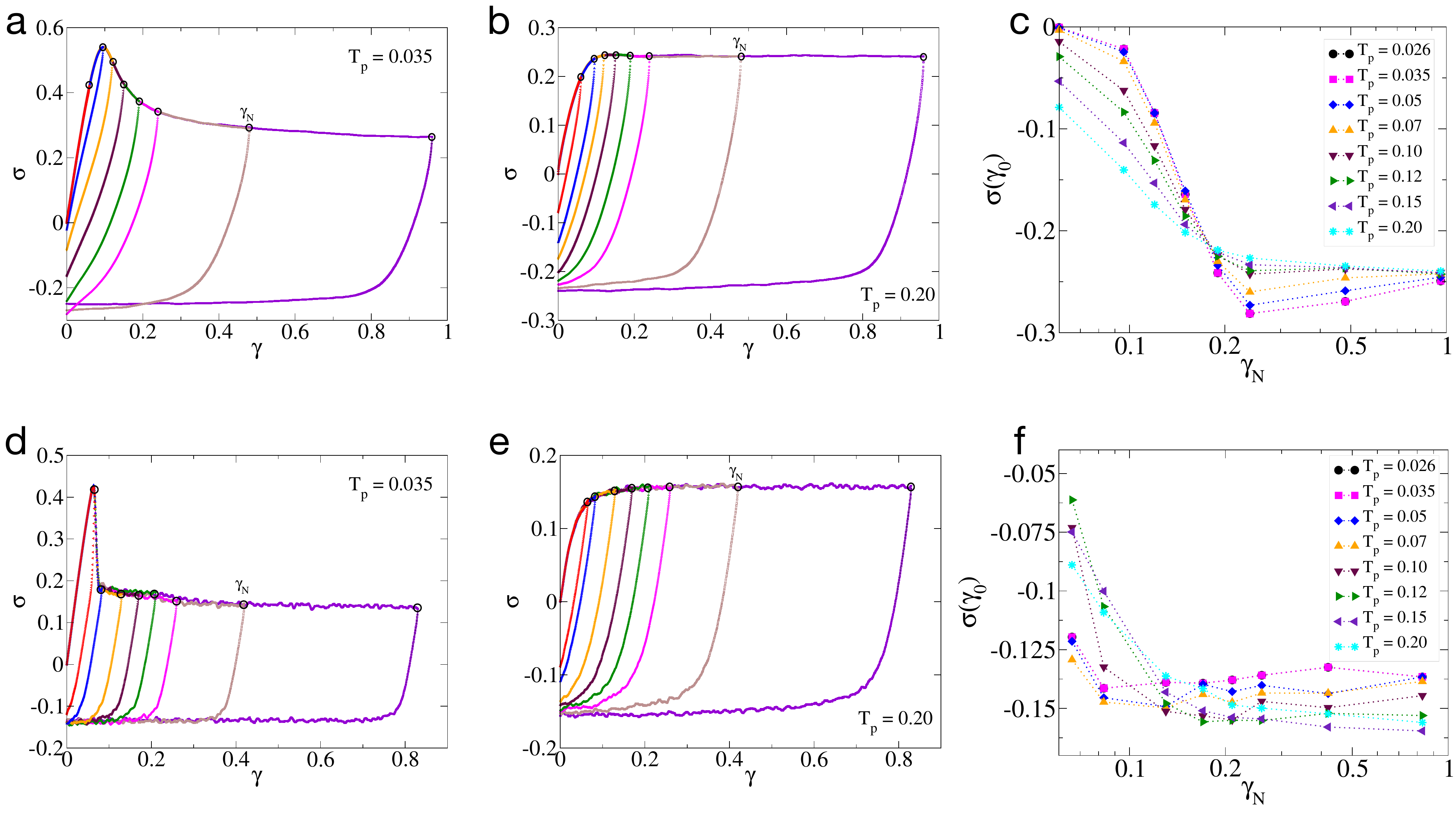}
\caption{Stress-strain curves for a system ($N=64{,}000$) sheared in the forward direction to different $\gamma_N$ values (black circles), and then reversed until $\gamma=0$. Panel (a) shows a well-annealed sample ($T_p=0.035$) at $\dot{\gamma}=5\times10^{-3}$ and panel (d) at $\dot{\gamma}=5\times10^{-5}$. Panel (b) shows a poorly annealed sample ($T_p=0.20$) at $\dot{\gamma}=5\times10^{-3}$ and panel (e) at $\dot{\gamma}=5\times10^{-5}$. Panels (c) and (f) show the residual stress at zero strain after a forward-reverse shear cycle to a maximum strain $\gamma_N$, over a wide range of annealing temperatures, at fast ($\dot{\gamma}=5\times10^{-3}$) and slow ($\dot{\gamma}=5\times10^{-5}$) shear rates, respectively. At the fast rate, well-annealed samples develop a strong, tunable negative residual stress (directional bias), whereas poorly annealed glasses show a weaker evolution. At the slow rate, well-annealed samples yield via a single system-spanning shear band, which limits the buildup of such directional memory, while poorly annealed samples show a similarly weak evolution with $\gamma_N$.}
\label{residualstress}
\end{figure}

We study the residual stress in the two yielding modes by calculating the stress at zero strain after the system has been sheared up to a maximum strain $\gamma_N$ in the forward direction and then brought back to zero strain by shearing in the reverse direction. We show the results for unidirectional shear of a well-annealed sample ($N=64{,}000$) prepared at $T_p=0.035$ at strain rates $5\times10^{-3}$ and $5\times10^{-5}$ in Fig.~\ref{residualstress}(a) and (d), respectively. The corresponding results for a poorly annealed sample prepared at $T_p=0.2$ sheared at $5\times10^{-3}$ and $5\times10^{-5}$ are shown in Fig.~\ref{residualstress}(b) and (e). 
For each state, we use different deformation histories, i.e. different values of $\gamma_N$, which lead to different residual stresses ($\sigma (\gamma_0)$) at $\gamma=0$. In Fig.~\ref{residualstress}(a,d), only the purely elastic regime at very small $\gamma_N$ traces back to zero stress upon reversal. In contrast, for the poorly annealed sample in Fig.~\ref{residualstress}(b,e), even small-amplitude loops do not return exactly to zero stress, because the high preparation temperature $T_p$ leads to enhanced thermally activated (viscoelastic) stress relaxation and the response is not strictly elastic.

For the well-annealed sample in Fig.~\ref{residualstress}(a), the residual stress decreases systematically with increasing $\gamma_N$, starting from a value close to zero, becoming increasingly negative, shows a slight reduction in magnitude, and then saturates as the steady state is approached. A similar trend is seen in Fig.~\ref{residualstress}(b) and (e) for the poorly annealed glass, but there the residual stress is already weakly negative at small $\gamma_N$, and the additional change with increasing $\gamma_N$ is smaller. However, this behavior is not observed in Fig.~\ref{residualstress}(d) for the well-annealed sample sheared at $5\times10^{-5}$. At this slower strain rate, the system yields abruptly via the formation of a single system-spanning shear band, which leads to a discontinuous drop in both the stress and the residual stress; as a result, the residual stress remains less negative than at the faster strain rate for comparable $\gamma_N$. 

In Fig.~\ref{residualstress}(c), the residual stress at zero strain, $\sigma(\gamma_0)$, is shown for different $T_p$ at the faster strain rate, and in Fig.~\ref{residualstress}(f) for the slower rate. Fig.~\ref{residualstress}(c) shows that for low $T_p$ the initial residual stress before yield is close to zero, and then becomes more negative after yielding before saturating. In contrast, at high $T_p$ the residual stress is already non-zero before yield, and $\sigma (\gamma_0)$ decreases more gradually and almost continuously with increasing $\gamma_N$ due to shear-induced rearrangements and ongoing viscoelastic relaxation. Thus, the imprint of shear on the residual stress, understood as the change from an initially vanishing $\sigma (\gamma_0)$ to a more negative value, is sharper and larger in well-annealed samples, whereas in poorly annealed samples the system starts from an already negative residual stress and the additional change with $\gamma_N$ is comparatively weaker. 

Taken together, these trends already contain a clear signature of directional memory. Residual stress at zero strain shows the directional bias in the system because of previous shear. The more negative $\sigma (\gamma_0)$ becomes with increasing $\gamma_N$, the stronger the directional bias is. At low $T_p$ and fast shear, the residual stress evolves from $\sigma (\gamma_0) \approx 0$ to large negative values as $\gamma_N$ increases, indicating a strong, tunable bias imprinted by the last shear direction. In contrast, high-$T_p$ glasses exhibit only a weak additional change of $\sigma (\gamma_0)$ with $\gamma_N$, consistent with a smeared-out, viscous response, while the abrupt, single-band yielding at slow rates in well-annealed samples or thermal relaxation in poorly-annealed samples limits how negative $\sigma (\gamma_0)$ becomes. Thus, the behavior of $\sigma (\gamma_0)$ already anticipates the contrasting directional memory in the brittle and ductile regimes that we quantify in the next section.

\section{Transition from Inverse to Classical Bauschinger Effect}

\begin{figure}[htb]
\centering
\includegraphics[width=0.95\textwidth]{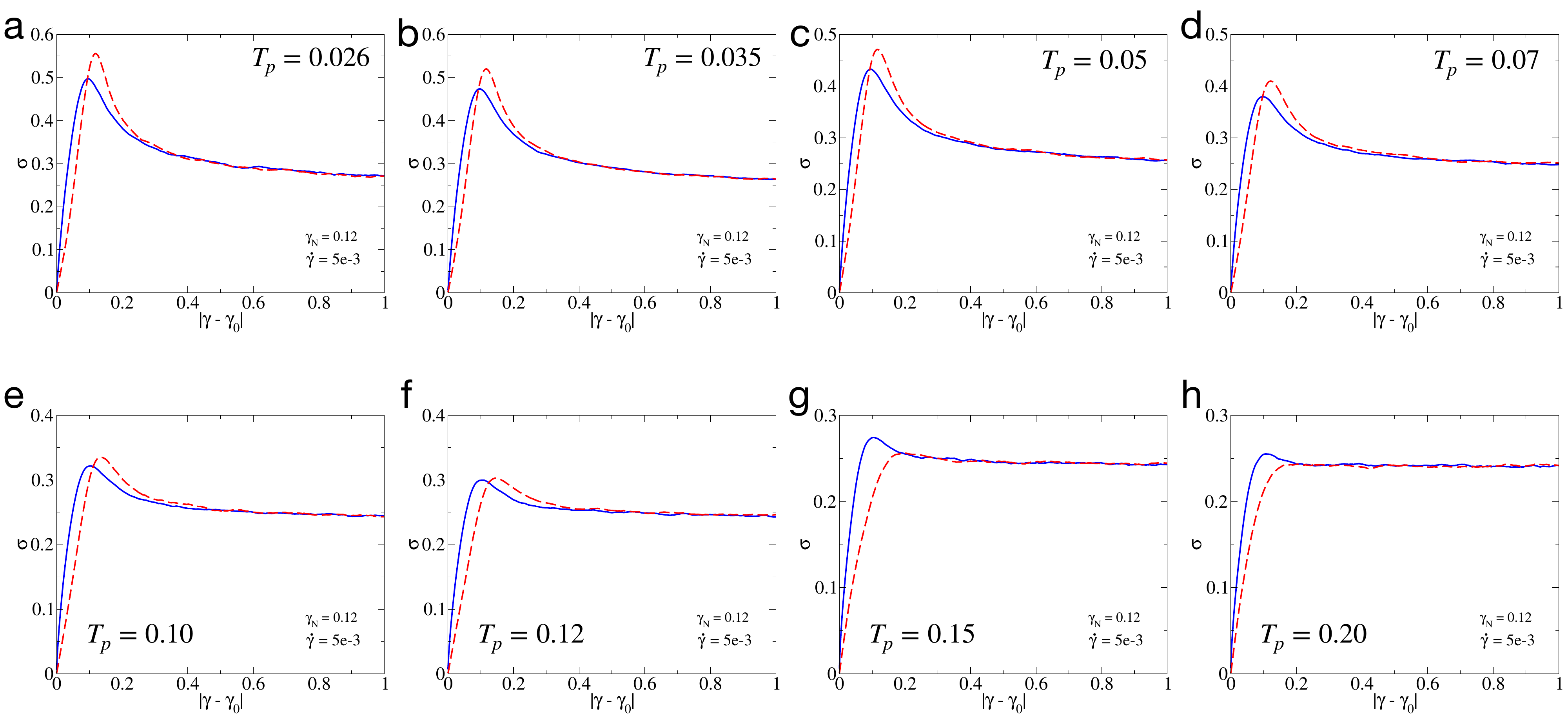}
\caption{Stress-strain curves showing the asymmetry between forward and reverse shear for samples with a pre-imposed directional bias, set by the maximum strain $\gamma_N=0.12$ to which they were previously sheared, while varying the preparation temperature $T_p$. Panels (a-h) show that, with decreasing annealing (increasing $T_p$) at $\dot{\gamma}=5\times10^{-3}$, the yield stress in the negative direction (red dashed, $\sigma_Y(n)$) changes from being higher than that in the positive direction (blue solid, $\sigma_Y(p)$) to being lower.}
\label{SM_Annealing}
\end{figure}
\begin{figure}[htb]
\centering
\includegraphics[width=0.95\textwidth]{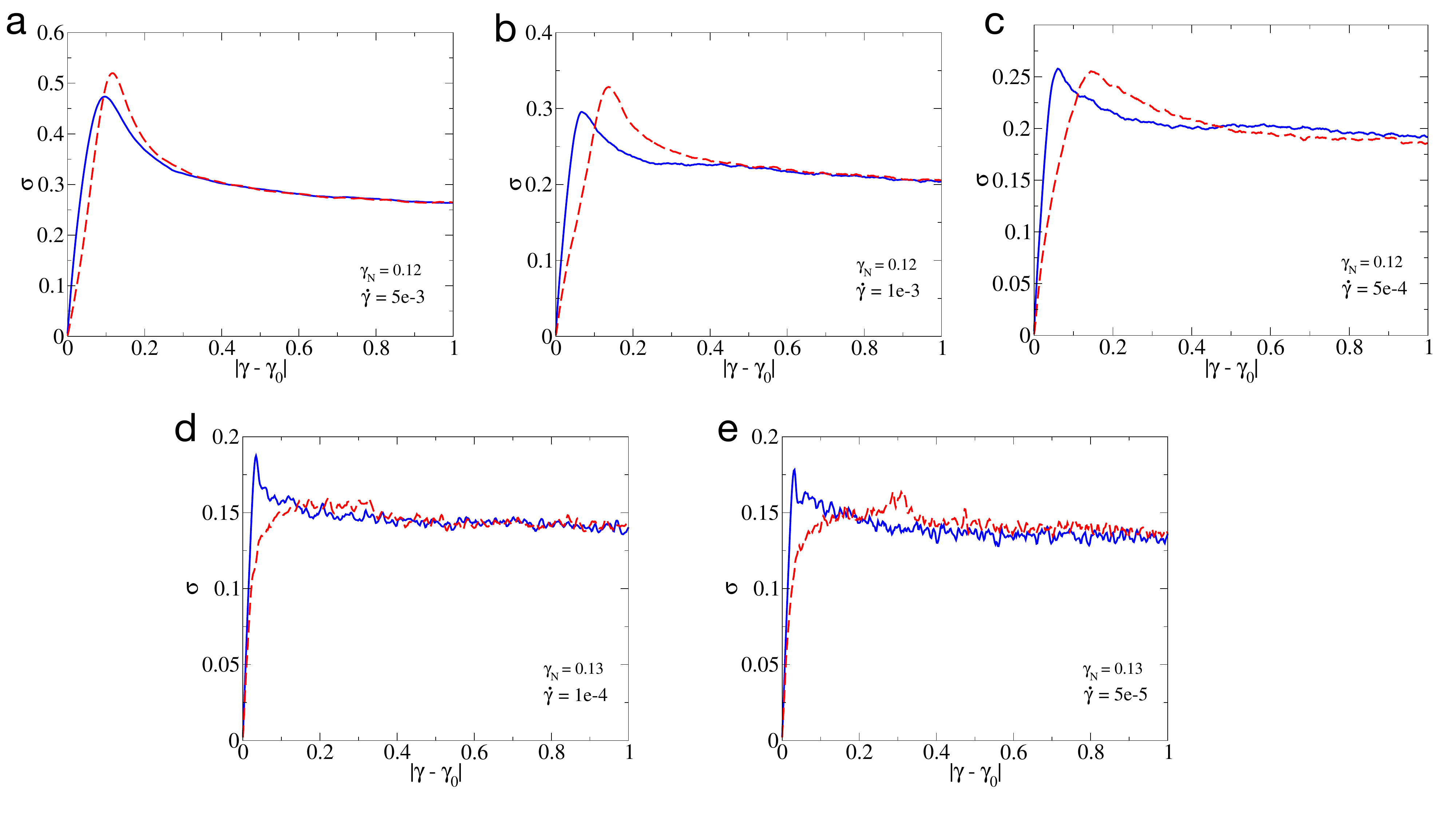}
\caption{Panels (a-e) show stress-strain curves for a well-annealed glass ($T_p=0.035$), illustrating the asymmetry between forward and reverse shear for samples with a pre-imposed directional bias set by the maximum strain $\gamma_N$, for decreasing strain rate. As the strain rate is reduced, the system transitions from IBE to BE, with the yield stress in the negative direction changing from being higher than that in the positive direction to being lower.}
\label{SM_strainrate}
\end{figure}

The transition from inverse to Bauschinger effect is shown by the stress-strain curves in Fig.~\ref{SM_Annealing}(a-h) as $T_p$ increases from $0.026$ to $0.20$. Each sample is first sheared to a maximum strain $\gamma_N = 0.12$, chosen near yield strain, and the deformation is applied at a strain rate $\dot{\gamma}=5\times10^{-3}$. Initially, the yield stress is larger in the reverse direction (red), showing an inverse Bauschinger effect, but this asymmetry decreases systematically with increasing $T_p$, becomes nearly symmetric around $T_p \approx 0.12$, and then crosses over to the conventional Bauschinger effect, where the yield stress in the forward direction becomes larger than in the reverse direction. But if one increases $\gamma_N$, the transition from the inverse to the classical Bauschinger effect shifts to lower $T_p$, since larger $\gamma_N$ drives the system deeper into the post-yield regime, where the shear-band morphology changes and the mechanism underlying the IBE is lost.

The transition from inverse to Bauschinger effect can also be realised by varying the shear rate, as shown in Fig.~\ref{SM_strainrate}(a-e) for a sample prepared at $T_p=0.035$. The stress-strain curves are plotted for strain rates ranging from $5\times10^{-3}$ to $5\times10^{-5}$. Each sample is first sheared to a maximum strain $\gamma_N = 0.12$ or $0.13$. At high strain rate, the yield stress is larger in the reverse direction (red) than in the forward direction (blue), corresponding to an inverse Bauschinger effect. As $\dot{\gamma}$ decreases, this asymmetry is progressively reduced, becomes nearly symmetric at an intermediate rate ($\dot{\gamma} \approx 5\times10^{-4}$), and eventually reverses, with the forward yield stress exceeding the reverse one, signalling the conventional Bauschinger effect.

\begin{figure}[htb]
\centering
\includegraphics[width=0.99\textwidth]{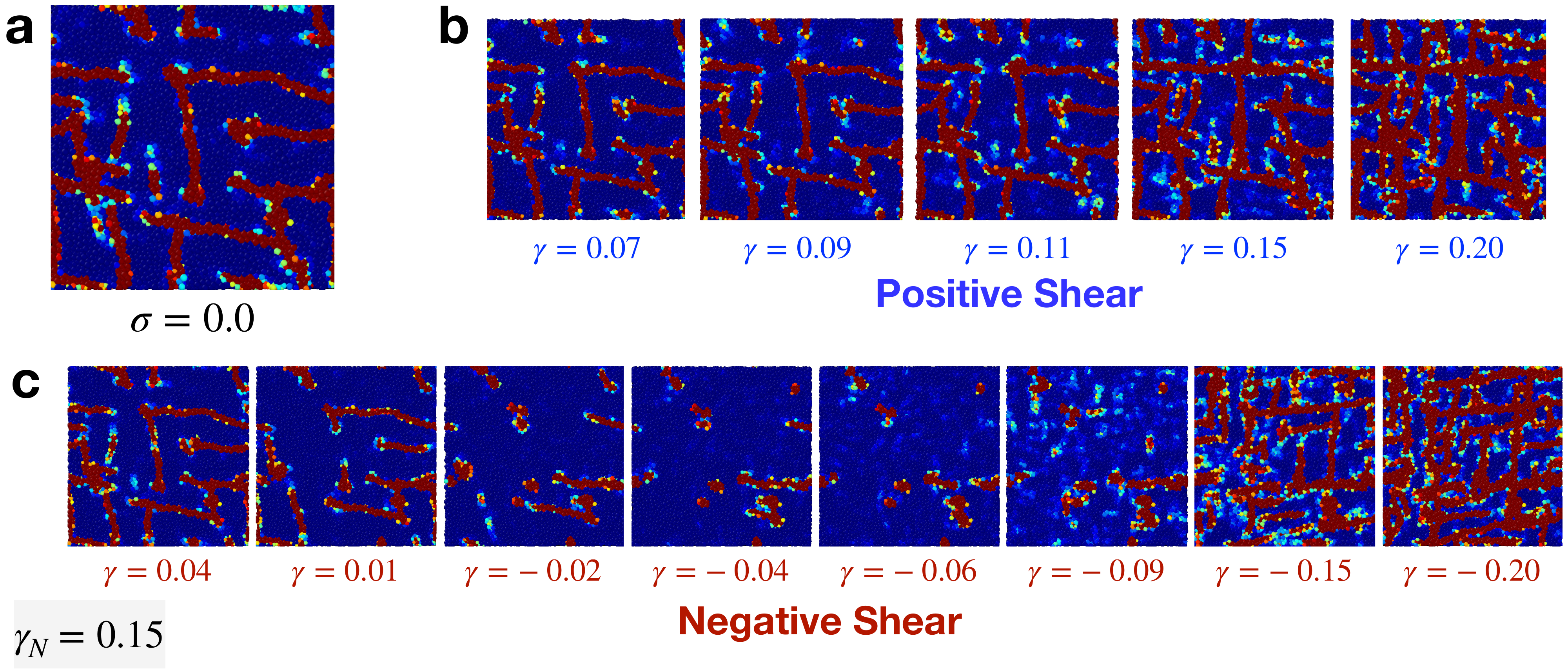}
\caption{In this figure, we observe the healing of the shear-band network when the system is sheared from a zero-stress state (panel (a), $\gamma=0.053250$) containing significant non-affine regions ($T_p=0.026$ and $N=10000$). Upon reversing the shear direction in panel (c), these bands begin to heal, but new shear-band networks nucleate again as the system approaches the yield point around $\gamma=-0.09$. In contrast, when sheared in the positive direction (panel (b)), the system continues to evolve the already formed shear bands without any noticeable healing.}
\label{HealingSB_sr5e-3}
\end{figure}
The same mechanism of plastic healing holds regardless of system size, as also seen for a smaller system ($N=10{,}000$) in a well-annealed glass prepared at $T_p=0.026$. The zero-stress state in Fig.~\ref{HealingSB_sr5e-3}(a) corresponds to this well-annealed sample sheared at $\dot{\gamma}=5\times10^{-3}$ up to $\gamma_N=0.15$, reaching a zero-stress state at strain $\gamma=0.05325$, where a network of shear bands has formed. Restarting shear from this state in the positive and negative directions leads to distinct evolutions of the shear-band pattern. Figure~\ref{HealingSB_sr5e-3}(c) shows the microscopic plastic-healing mechanism in the negative direction for $N=10{,}000$, provided a multiple shear-band network is present. In contrast, no comparable healing occurs in the positive shear direction (Fig.~\ref{HealingSB_sr5e-3}(b)).